\documentclass[prx,twocolumn,amsmath,amssymb,superscriptaddress,longbibliography,floatfix]{revtex4-2}
\usepackage[T1]{fontenc} 
\usepackage[english]{babel}
\usepackage{graphicx}
\usepackage{dcolumn}
\usepackage{amsbsy}
\usepackage{amstext}
\usepackage{amsmath}
\usepackage{amsfonts}
\usepackage{bm}
\usepackage{comment}
\usepackage[dvipsnames]{xcolor}
\usepackage[normalem]{ulem}

\usepackage[final]{hyperref} % adds hyper links inside the generated pdf file
\hypersetup{
	colorlinks=true,       % false: boxed links; true: colored links
	linkcolor=blue,        % color of internal links
	citecolor=blue,        % color of links to bibliography
	filecolor=magenta,     % color of file links
	urlcolor=blue         
}

\begin{document}
\title{Angular-momentum-selective nanofocusing with Weyl semimetals}
\author{Marco Peluso}
    \affiliation{Dipartimento di Scienza Applicata e Tecnologia, \\ Politecnico di Torino, Corso Duca degli Abruzzi 24, 10129 Torino, Italy}
\author{Alessandro De Martino}
    \affiliation{Department of Mathematics, City St~George's, University of London,
    Northampton Square, EC1V OHB London, United Kingdom}
\author{Reinhold Egger}
    \affiliation{Institut f\"ur Theoretische Physik,
    Heinrich-Heine-Universit\"at, D-40225  D\"usseldorf, Germany}
\author{Francesco Buccheri}
    \affiliation{Dipartimento di Scienza Applicata e Tecnologia, \\ Politecnico di Torino, Corso Duca degli Abruzzi 24, 10129 Torino, Italy}
    \affiliation{INFN Sezione di Torino, Via P. Giuria 1, I-10125, Torino, Italy}
\date{\today}
\begin{abstract}
    We investigate the theory of surface plasmon polaritons on a magnetic Weyl semimetal conical tip. We show that the axion term in the effective electrodynamics modifies the surface plasmon polariton dispersion relation and allows all modes with a given sign of the orbital angular momentum to be focused at the end of the tip. This is in contrast with normal metals, in which only one mode can reach the end. We discuss how this orbital angular momentum nanofocusing expands the potential of technologies that use this degree of freedom.
\end{abstract}
\maketitle
\section{Introduction}
The phase modulation of a beam wavefront is encoded in its orbital angular momentum (OAM) number, which takes arbitrary integer values. In that, it is distinct from the "spin" angular momentum, which describes the binary polarization of the light beam \cite{Allen1992,RubinszteinDunlop2017,Babiker2019}. 
OAM is an important resource in present-day and future technologies. For instance, quantum communication protocols are essentially based on photons, which can transmit information over long distances \cite{Yin2017}. 
In this respect, the OAM significantly improves the amount of information that can be shared between an emitter and a receiver, because it is associated with an infinitely-dimensional discrete Hilbert space. The OAM multiplexing is an example of the more efficient use of communication channels allowed by this degree of freedom \cite{Gibson2004,Yue2018,Zahidy2022}. The improved dimensionality also allows more secure communications, with the implementation of advanced quantum cryptography protocols \cite{Zahidy2024}.
Other fields of application include nanophotonic devices \cite{Jia2016}, holography \cite{Kong2023}, quantum \cite{GarciaEscartin2011} and classical computation, e.g., specialized machine learning algorithms \cite{Fang2024}.
Experimentally, the continued attention and the mature technologies have achieved important goals, e.g., multidimensional entanglement between OAM-carrying photons \cite{Mair2001} and the quantum teleportation of this degree of freedom, in addition to the spin angular momentum \cite{Wang2015}. 
Twisted light can be used to trap and manipulate small particles and atoms %and molecules 
\cite{Torres2011,Maslov2024,Babiker2019}, which has been observed in various configurations \cite{Simpson1997,He1995,Oneil2002}. Due to the possibility of selectively eliciting transitions between discrete states, OAM is also a powerful spectroscopic tool and allows, for example, the optical resolution of the different enantiomers of a chiral molecule \cite{Bradshaw2015,Brullot2016}, as well as complex object assessment \cite{Cheng2025}.
\begin{figure}[h]
    \centering
    \includegraphics[width=1\columnwidth]{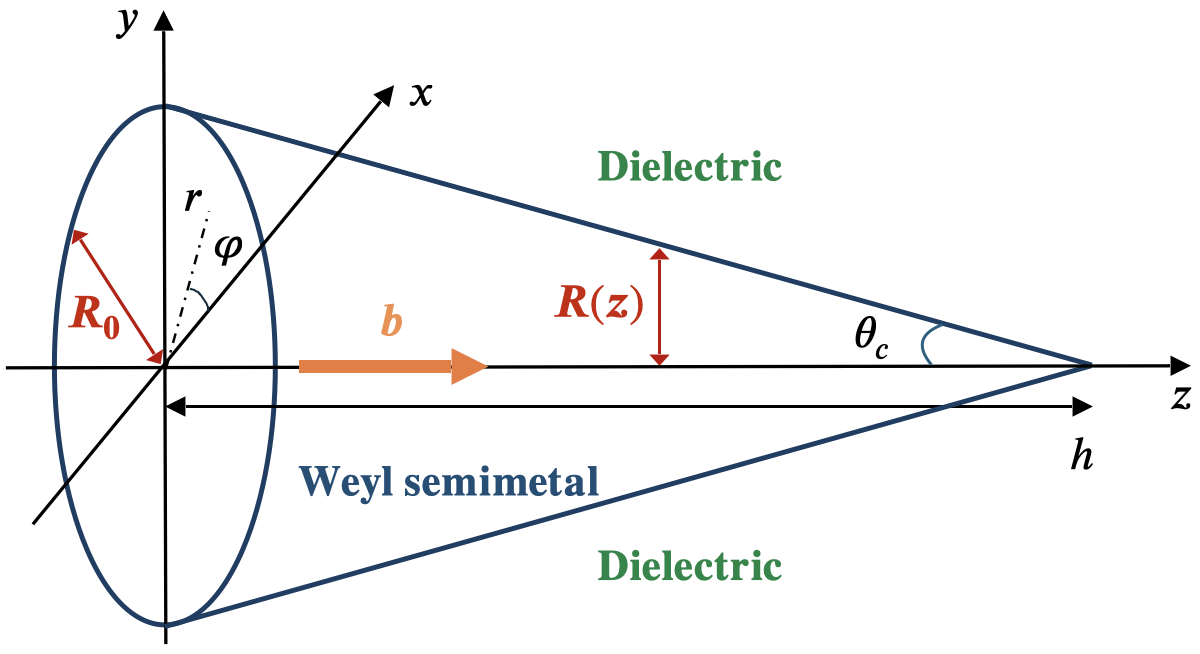}
    \caption{Conical waveguide scheme: a magnetic Weyl semimetal core surrounded by a dielectric medium. 
    The vector $\mathbf{b}$ oriented along the $z$ direction describes the Weyl node separation in momentum space. 
    The cone has an opening angle $\theta_c$. For further details, see main text. }
    \label{fig:cone}
\end{figure}
For all these reasons, control of the OAM at micro- and nanoscopic scale is highly desirable, as it would open possibilities to precisely manipulate nanostructures, and offer advantages in terms of spectroscopic resolution, increased power efficiency, downsizing and scaling of quantum technologies. 

However, the possibility of miniaturizing photonic devices is fundamentally limited by diffraction, which prevents focusing of light beams by purely optical means. One route to overcome this limitation is normally offered by surface plasmon polaritons (SPPs) \cite{Yu2019}, coherent excitations of light and matter localized at the interface between a metallic state and a dielectric. SPPs offer a unique combination of properties, such as high spatial confinement and strong field enhancement \cite{Chang2006}. 
 Due to the sub-wavelength confinement, SPPs allow the reduction of the size of photonic devices down to nanometric scale \cite{Oulton2009,Gramotnev2010,Stockman11}. 
 Because of the strong interaction of light and matter due to the field enhancement, they find a wide range of applications, see e.g., \cite{Yu2019,Patching2014}.
 In conical structures, SPPs are used to improve the resolution of near-field optical microscopy and scanning tunneling microscopy, allowing for ultra-local, non-destructive probes \cite{Mauser2005,DeAngelis2010,DeAngelis2011,Das2016}. While the SPP nanofocusing has been studied by theoretical and experimental means before \cite{Stockman04,Wiener2012,Choi2009}, the possibility of spatially concentrating OAM-carrying radiation seems precluded so far \cite{Chang2007}.

In this work, we theoretically investigate the OAM nanofocusing by conical tips, see Fig. \ref{fig:cone} for an illustration, and we show that this limitation can be overcome. In particular, we show that, while OAM beams cannot be spatially compressed in ordinary metals below a certain radius, gyrotropic materials can focus it down to nanoscopic scale. We analyse the class of Weyl semimetals (WS), which exhibit pairs of topologically protected band crossings, or Weyl nodes \cite{Wan2011,Hosur2012,Armitage2018} and a simple, (quasi-)universal and especially large static limit of the off-diagonal permittivity. The bulk band structure, as well as the surface states and the surface transport, are robust against weak interactions, e.g., with impurities or phonons \cite{Buchhold2018,Yang2019,Buccheri2022}, which makes them promising candidates for electronic technologies. We focus on materials with broken time-reversal symmetry, or magnetic WSs, in which the presence of Weyl nodes is at the origin of the remarkable optical activity  \cite{Kargarian2015,Chen2019}. Indeed, the bulk topology manifests itself in the appearance of an axionic $\theta$-term in the effective electrodynamics \cite{Zyuzin2012,Vazifeh2013,Chen2013}, originating the giant anomalous Hall effect \cite{Liu2018}, and of topological magnetoelectric and magnetotransport effects \cite{Vanderbilt2018,Sekine2021}. Importantly, the $\theta$-term determines a nontrivial SPP dynamics at planar interfaces and cylindrical waveguides \cite{Song2017, Andolina2018, Weylguide2024} and modifies the spectrum of localized surface plasmons \cite{Pellegrino2025}.

We show that the OAM nanofocusing in topological WSs can be described by a single parameter, which can be independently determined, e.g., from transport measurements. The origin of this parameter is in material-independent low-energy properties of the electronic bands.
Depending on the material orientation, modes with one sign of the OAM are always emitted, while all the others always reach the tip. 
We determine the frequency range over which the phenomenon occurs as a function of the plasma frequency and the Weyl node separation.

The paper is structured as follows. In Section \ref{sec:Model}, we introduce the model of WS employed in this study and we review the necessary elements about SPPs in cylindrical waveguides. Sec. \ref{sec:cone} introduces the conical geometry under consideration and the necessary formalism. Sec. \ref{sec:nanofocusing} demonstrates the occurrence of the OAM nanofocusing in WSs, adapts the model to materials with multiple Weyl nodes and provides results for a case study. We add a discussion of potential impact and applications in Sec. \ref{sec:discussion}. Technical details of our calculations can be found in the Appendices.
\section{Model}
\label{sec:Model}
In this work, we study a WS conical tip embedded in a dielectric environment. 
The WS features a single pair of Weyl nodes, labelled by their chirality $\chi=\pm1$. Near the crossing points, the band structure of the bulk three-dimensional material is described by the low-energy Hamiltonian \cite{Burkov2011}
\begin{equation}\label{eq:Hchi}
    H_\chi(\mathbf{k})=\hbar v_F \bm{\sigma} \cdot \left( {\mathbf k} - \chi {\mathbf b}\right),
\end{equation}
where $\bm{\sigma}=(\sigma_x,\sigma_y,\sigma_z)$ are the Pauli matrices, $\mathbf{k}$ the electronic crystal momentum, and $v_F$ is the Fermi velocity. We denote the vector that separates the Weyl nodes as $2\mathbf{b}$ and set its orientation along the z direction. The robustness of the WS phase can be formalized by defining a topological charge associated to each node, as the flux of the Berry curvature across a sphere surrounding it \cite{Armitage2018}. As a consequence, Weyl nodes can only be annihilated in pairs \cite{Nielsen1981b,Nielsen1981c}, which makes the topological phase robust against weak disorder \cite{Buchhold2018}.

In the WS, the electric displacement $\mathbf{D}=\hat{\varepsilon}\mathbf{E}$ 
is related to the electric field via the permittivity tensor $\hat{\varepsilon}$. 
The latter is determined from the conductivity $\hat{\sigma}$ through the relations
\begin{equation}   \label{eq:epstosigma}
    \hat{\varepsilon}=\varepsilon_0\epsilon_W+\frac{i\hat\sigma}{\omega} = \varepsilon_0 \mathcal{E}\mathbb{I}+\frac{i \sigma_H\hat{e}_H}{\omega} \,.
\end{equation}
Here we have denoted $\varepsilon_0$ the permittivity of the vacuum, $\epsilon_W$ the relative background permittivity, $\mathcal{E}=\mathcal{E}(\omega)$ the diagonal part of the relative permittivity, $\hat{e}_H$ the antisymmetric tensor with $1$ in the $yx$ component and
\begin{equation} \label{eq:sigmaH}
    \sigma_H=\frac{e^2 b}{\pi h} %\frac{2\alpha \varepsilon_0 c b}{\pi}=
\end{equation}
the Hall conductivity \cite{Burkov2014}. Throughout this paper, we assume a simple Drude form in the dissipationless limit of the diagonal conductivity
\begin{equation}
    \mathcal{E}(\omega)=\epsilon_{W}\left[1-\left(\frac{\omega_p}{\omega}\right)^2\right] \;, 
\end{equation}
where $\omega_p$ is the plasma frequency, see Sec. \ref{sec:multiple} for numerical values, 
and ignore non-local and temperature corrections. Inter-band transitions do not affect strongly this function below the plasma frequency \cite{Weylguide2024}.
 The electric displacement in the dielectric is $\mathbf{D}=\varepsilon_0\epsilon_d\mathbf{E}$, with $\epsilon_d$ the relative permittivity. 
 Finally, we assume that in both materials, the magnetic induction field is proportional to the magnetic field with unit relative magnetic permeability $\mathbf{B}=\mu_0\mathbf{H}$.

\subsection{The WS cylindrical waveguide}\label{sec:cylinder}
\begin{figure}[h!]
    \centering
    \includegraphics[width=0.49\columnwidth]{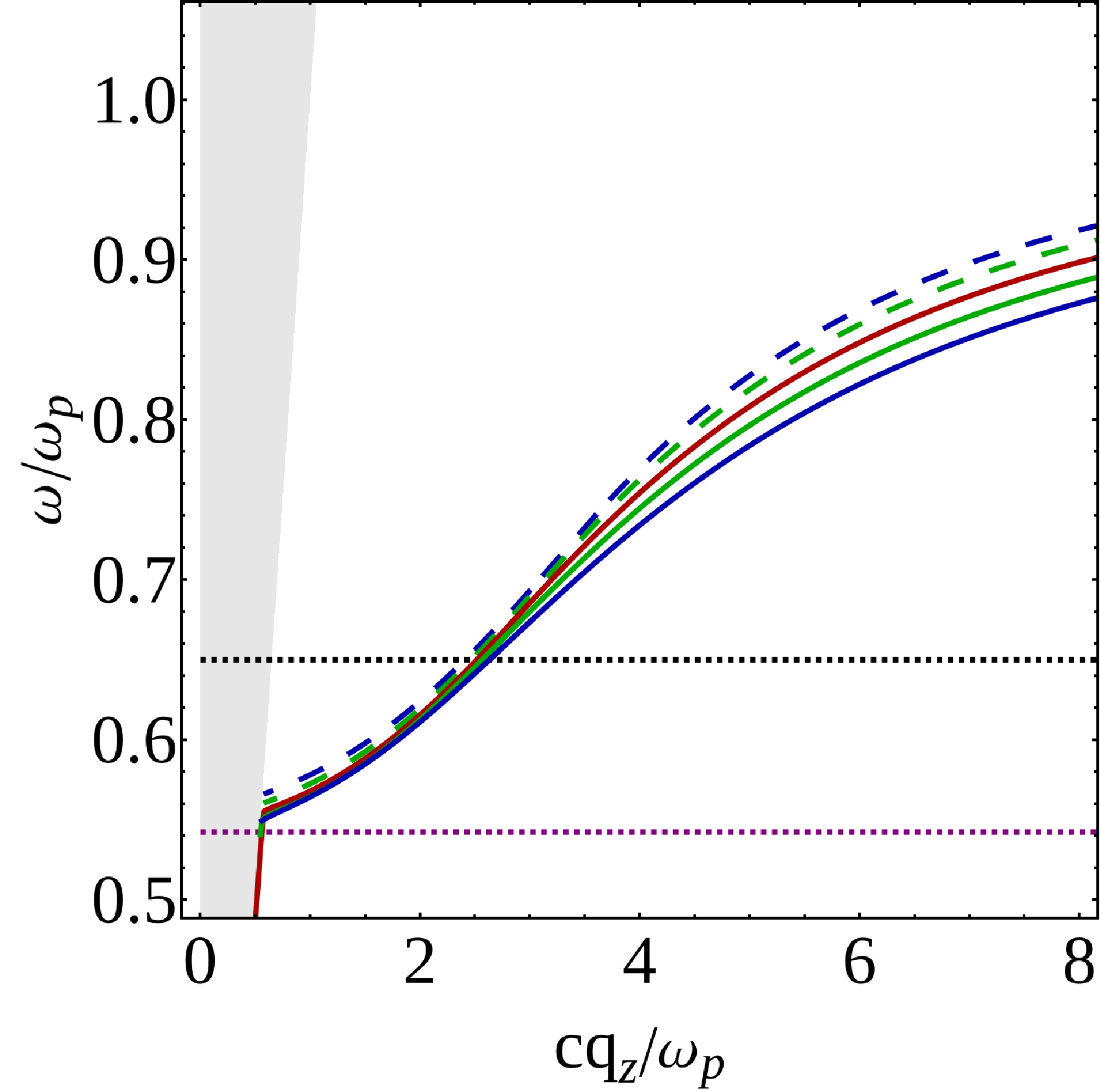}
        \includegraphics[width=0.49\columnwidth]{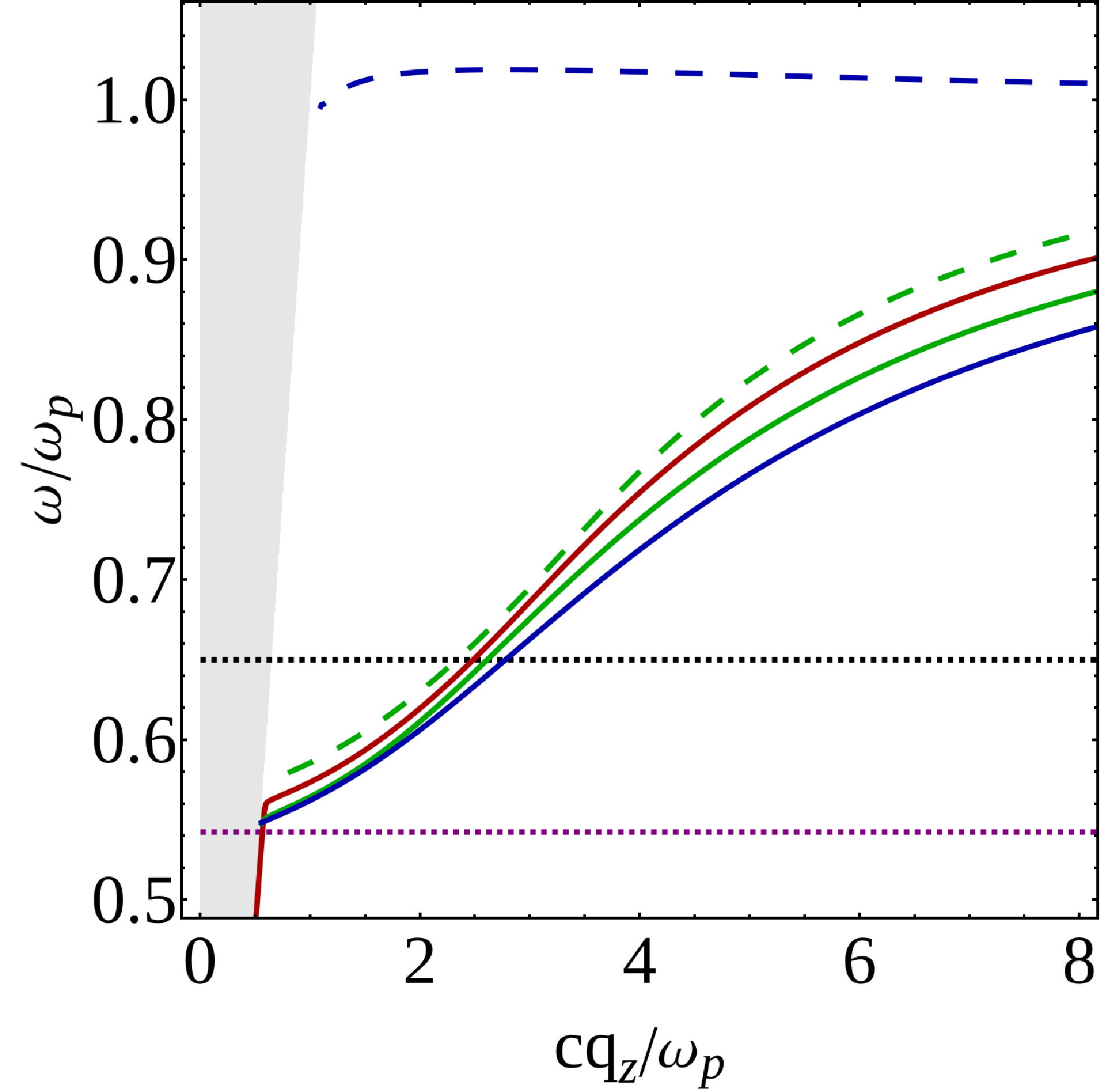}
        \caption{Size dependence of the SPP dispersion relation in a cylindrical WS waveguide. The parameters were chosen as $\beta=16$, $\epsilon_W=25$, $\epsilon_d=1$ and $\rho=4$ (left), $\rho=2.4$ (right). Here we show the bands $m=0$ (solid red), $m=1$ (dashed green), $m=-1$ (solid green), $m=2$ (dashed blue) and $m=-2$ (solid blue). The dotted purple line is the frequency $\omega_-$, see Eq. \eqref{eq:omega-}. The dotted black line denotes a given frequency, here $\omega=0.65\omega_p$. For the larger section, the waveguide supports the $m=2$ mode at that frequency, while for the smaller one, this mode is not supported.}
    \label{fig:cylwaveguideRcyl1}
\end{figure}
In this section, we briefly review the SPP modes supported by a cylindrical WS of radius $R$  \cite{Weylguide2024}, 
with emphasis on the radius dependence of the modes. 
In cylindrical geometry, the rotational symmetry around the axial direction $z$ allows us to expand the fields as
\begin{equation}\label{eq:angmombasis}
    \mathbf{E}(\mathbf{r},z)=\sum_{m\in\mathbb{Z}}\mathit{e}^{\mathrm{i}m\varphi}\mathbf{E}_m(r,z)\,,
\end{equation}
where $m$ is the OAM and $r=|\mathbf{r}|$ is the radial coordinate, with $\mathbf{r}=(x,y)$. Using also the translational invariance along $z$, the field components can be factorized as $\mathbf{E}_m(r,z)=\widetilde{\mathbf{E}}_m(r)\mathit{e}^{\mathrm{i}q_zz}$, where $q_z$ is the wave number along $z$. 
The modes  $\widetilde{\mathbf{E}}_m(r)=\mathbf{E}_{m,W}\Theta\left(R-r \right)+\mathbf{E}_{m,d}\Theta\left(r-R\right)$ are solutions of the electrodynamics equations \eqref{eq:GaussE}-\eqref{eq:AmpMax} and are written explicitly in Appendix \ref{app:polarization_vectors}. We are interested in the SPP  solutions, which are exponentially localized around the interface between the WS and the dielectric \cite{Weylguide2024}.
They are characterized by their inverse localization lengths
\begin{equation}\label{eq:kappa_diel}
    \frac{\kappa_d}{k_0}=\sqrt{n^2-\epsilon_d}
\end{equation}
in the dielectric, and
\begin{align}\label{eq:kappagamma}
    \frac{\kappa_\pm}{k_0}&=\sqrt{n^2-\mathcal{E}+2\eta\gamma_\pm}, &
    \gamma_\pm&=\frac{\mathcal{\eta}}{\mathcal{E}}\pm\sqrt{\frac{\eta^2+n^2\mathcal{E}}{\mathcal{E}^2}}
\end{align}
in the WS, where $k_0=\omega/c$, $n(\omega)=cq_z/\omega$ is the refractive index
and $\eta\left(\omega\right)={2\alpha c b}/{\pi\omega}$, with $\alpha\approx 1/137$ the fine structure constant.
The SPP dispersion relation can be obtained by imposing continuity conditions for the tangential components of the electric field and the magnetic field
\begin{align}
\label{eq:BoundCond}
    \mathbf{E}_{m,W}^{\parallel}(R)=\mathbf{E}_{m,d}^{\parallel}(R)&, &
    \mathbf{H}_{m,W}^{\parallel}(R)=\mathbf{H}_{m,d}^{\parallel}(R)&\,.
\end{align}
This leads to a linear homogeneous system, which only has solutions if the secular equation
\begin{equation}\label{eq:SecularEquation}
    \det{B_m}=0
\end{equation} 
is satisfied, for the matrix
\begin{equation}\label{eq:MatrixBoundCondMain}
B_m=
    \begin{bmatrix}
        u_+ &
        u_- &
        v & 
        0 \\
        \frac{u_+}{\gamma_+}\frac{n^2}{\mathcal{E}} &
        \frac{u_-}{\gamma_-}\frac{n^2}{\mathcal{E}} &
        0 &
        v \\
        \frac{\gamma_+}{u_+}m+\frac{I_{m}'(u_+)}{I_{m}(u_+)} &
        \frac{\gamma_-}{u_-}m+\frac{I_{m}'(u_-)}{I_{m}(u_-)} &
        \frac{K_{m}'(v)}{K_{m}(v)} &
        \frac{m}{v} \\
        \frac{1}{\gamma_+}\frac{I_{m}'(u_+)}{I_{m}(u_+)}+\frac{m}{u_+} &
        \frac{1}{\gamma_-}\frac{I_{m}'(u_-)}{I_{m}(u_-)}+\frac{m}{u_-} & 
        \frac{m}{v} &
        \frac{\varepsilon_d}{n^2}\frac{K_{m}'(v)}{K_{m}(v)}
    \end{bmatrix} \,.
\end{equation}
Here, the modified Bessel functions $I_m$ and $K_m$ appear, with arguments $u_\pm= R\kappa_\pm$ and $v=R\kappa_d$, see \eqref{eq:kappagamma}.

The waveguide radius determines the form of the SPP dispersion curves $\omega_m\left(q_z\right)$ through the dimensionless parameter $\rho = \omega_p R/c$, together with the separation of the Weyl nodes encoded in the parameter 
\begin{equation}
    \beta = \frac{\alpha c b}{\pi\omega_p} \,,
\end{equation}
When $\beta > 0$, the modes with OAM $m$ and $-m$ are non-degenerate at all sizes. The splitting is qualitatively consistent with finite-size quantization, and is vanishingly small in the planar limit $\rho\to\infty$, while it increases for decreasing $\rho$.
This has far-reaching consequences for what the nanofocusing is concerned, which can be understood from general observations.
For given value of $\beta$, we shall be interested in the frequency range
\begin{equation}\label{eq:d}
    \omega_-<\omega<\omega_\infty\,,
\end{equation}
where
\begin{equation}\label{eq:omega-}
    \omega_-= \frac{\epsilon_W \omega_p}{\sqrt{\epsilon_W\left(\epsilon_W+\epsilon_d\right)+\beta^2}+\beta}
\end{equation}
is the $m\to-\infty$ asymptote and
\begin{equation}\label{eq:omegainfty}
    \omega_\infty=\omega_p\sqrt{\frac{\epsilon_W}{\epsilon_d+\epsilon_W}}\,
\end{equation}
is the large wavevector asymptote \cite{Weylguide2024}.
At very large radii, at any given frequency in the above range, both modes with $m$ and $-m$ appear, although at different wavevectors.
Conversely, at smaller radii, the dispersion of the modes with $m>0$ is shifted to higher frequencies. 
As a consequence, below a critical radius $R_m$, 
the waveguide will support only the SPP modes with negative OAM. The critical radii $R_m$ will be discussed in detail in Sec. \ref{sec:nanofocusing}. In order to illustrate this behavior, Figure \ref{fig:cylwaveguideRcyl1} displays the dispersion relations of few SPP modes with lowest OAM in two WS cylindrical waveguides with different radii.
At the frequency chosen in our example, both $m=\pm2$ modes are sustained by the waveguide with larger radius, see Fig. \ref{fig:cylwaveguideRcyl1}. 
Conversely, for the smaller radius, the $m=+2$ mode  is no longer allowed, while its counterpart $m=-2$ is. 
This is to be contrasted with the scenario of a normal metal ($\beta=0$), 
in which the two modes at fixed OAM remain degenerate at all sizes 
and only the $m=0$ mode is present for all values of $R$ \cite{Stockman04,Chang2007}.

\section{The conical tip}\label{sec:cone}
We now turn to the conical geometry, illustrated in Fig. \ref{fig:cone}. 
This is modelled by introducing a $z$-dependence in the radius for $0<z<h$, 
where $h$ is the height of the cone and the apex is located at $x=y=0$, $z=h$.
The radius is given by
\begin{equation}
    R \to R(z)=\left(1-\frac{z}{h}\right)R_0\;,
    \label{zdepradius}
\end{equation}
where $R_0$ is the base radius. In the figures below, as geometric parameters, we choose $h=50c/\omega_p$ and $R_0=2c/\omega_p$, see Fig. \ref{fig:cone}, which results in an opening angle $\theta_c=0.04$. However, the theory described below applies to arbitrary but small values of $\theta_c$. 
The rotational symmetry of the system around the $z$ axis remains unbroken and therefore we can use the expansion of the fields as in Eq. \eqref{eq:angmombasis}. 
While plane wave factorization is no longer possible, it is still possible to study analytically the problem if the opening angle of the cone is small. In this situation, we may write the electromagnetic field in WKB (adiabatic) approximation. In particular, the $z$-dependence of the field will be encoded in 
 the factorized amplitude $\mathcal{A}(z)$ and in its phase \cite{Stockman04},
\begin{equation}\label{eq:Epolvec}
    \mathbf{E}_m(r,z)=\widetilde{\mathbf{E}}_m(r,n_m)\mathcal{A}_m(z)\exp{\left\{\mathrm{i}\frac{\omega}{c}\int_{0}^z n_m(\omega,s)\mathrm{d}s\right\}} \;.
\end{equation}
Here $\widetilde{\mathbf{E}}_m(r,n_m)$ satisfies the Maxwell equations
 as long as the adiabatic parameter
\begin{equation}\label{eq:delta}
    \delta_m=\left|\frac{c}{\omega n_m^2}\frac{\partial n_m}{\partial z}\right|
\end{equation}
is much smaller than $1$ \cite{Stockman04}, which translates to a condition on the aperture of the cone \mbox{$R_0/h \ll 1$}. Details are provided in Appendix \ref{app:polarization_vectors}.
The continuity condition \eqref{eq:SecularEquation} determines the refractive index  
$n_m(\omega,z)$ as a function of the frequency and the local radius. Therefore, the vector $\widetilde{\mathbf{E}}_m$ implicitly depends on the position along the cone. The amplitude $\mathcal{A}$ is obtained by imposing the conservation of the flux of the Poynting vector through a plane perpendicular to the cone axis. This quantity, averaged over a period, is written as
\begin{equation}
\Phi_m = \pi\left|\mathcal{A}_m(z)\right|^2\int_0^\infty\operatorname{Re}\left\{\widetilde{\mathbf{E}}_m\times\widetilde{\mathbf{H}}_m^*\right\}_z r\mathrm{d}r\,,
\end{equation}
and is independent of the position along $z$.
The quantities of interest in our analysis will be the SPP phase and group velocities
\begin{align}\label{eq:vpvg}
    v_{p,m}&=\frac{c}{n_m}, &
    v_{g,m}&=c\left\{\frac{\partial \left[n_m\omega\right]}{\partial \omega}\right\}^{-1}\,.
\end{align}
The phase velocity changes adiabatically along the cone and, when it reaches the value $c/\sqrt{\epsilon_d}$, 
the SPP is emitted in the dielectric \cite{Chang2007}. Indeed, at the position in which the local 
index of refraction takes the value $n_m=\sqrt{\epsilon_d}$, the localization length diverges 
on the side of the dielectric, see Eq. \eqref{eq:kappa_diel}.
A useful quantity to characterize the nanostructure is the field intensity relative 
to a given reference position $(\mathbf{r}_0,z_0)$
\begin{equation}\label{eq:LFIE}
L\left(\omega,\mathbf{r},z\right)=\frac{\left|\mathbf{E}\left(\omega,\mathbf{r},z\right)\right|^2}
{\left|\mathbf{E}\left(\omega,\mathbf{r}_0,z_0\right)\right|^2}\,.
\end{equation}
This is known as local field intensity enhancement (LFIE) factor and expresses 
how the presence of the tip affects the electric field locally \cite{Maslovski2019,Stockman11}. 
The energy density can be directly computed from the field intensity and the group velocity
\cite{Stockman04,LandauLifshitzElectrodynamics}.
\begin{figure}[h]
    \centering
     \includegraphics[width=\columnwidth]{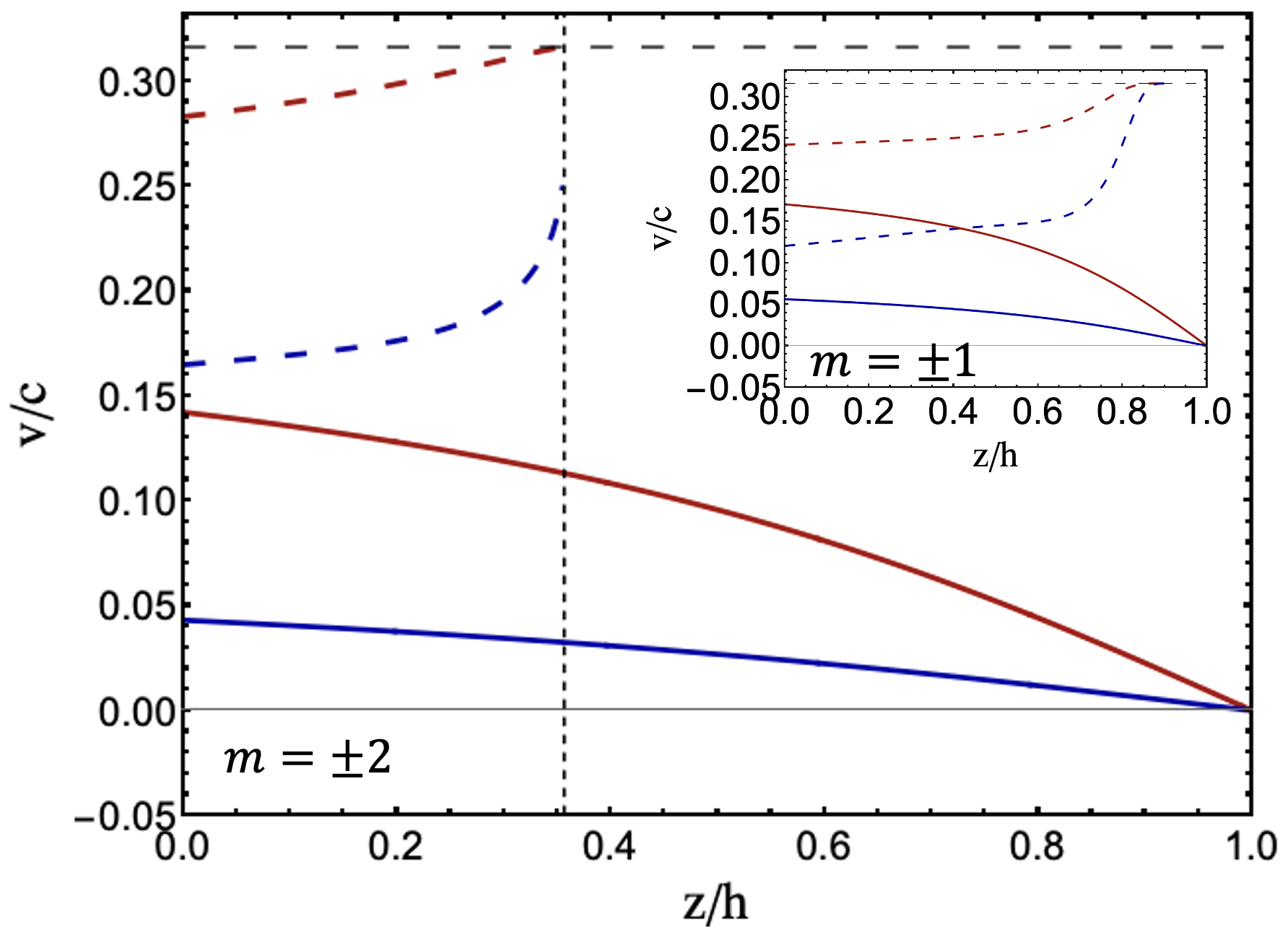}
    \caption{    
    Phase and group velocities \eqref{eq:vpvg} along the cone (red and blue curves, respectively).
    Main panel: modes with $m=+2$ (dashed) and $m=-2$ (solid). The blue dashed curve reaches 
    the horizontal dashed line.
    Inset: same for $m=\pm 1$. Dashed curves asymptotically approach the horizontal 
    dashed line, but numerical accuracy prevents resolution of their final segment near this asymptote.
    The parameters used here are: $\omega/\omega_p=0.5$, $\epsilon_d=10$, $\epsilon_W=10$, and $\beta=10$. 
    The $m=-2$ mode reaches the cone tip with vanishing phase and group velocities, 
    therefore focusing the electromagnetic field energy,     
    analogous to the  $m=0$ mode of a conventional metal. In contrast, 
    the $m=2$ mode is radiated into the dielectric at the radius 
    $R_m$ (see Sec. \ref{sec:nanofocusing}), indicated by the vertical dashed line. 
    The horizontal dashed line marks $c/\sqrt{\epsilon_d}$, the velocity at which the  $m=+2$ mode is radiated. We note that $\delta_m$ in Eq.~\eqref{eq:delta} remains very small throughout the shown range of parameters.
    }
    \label{fig:velocities}
\end{figure}
\section{OAM nanofocusing}\label{sec:nanofocusing}
As reviewed in Sec. \ref{sec:cylinder}, in a cylindrical waveguide, the dispersions of modes with the same absolute value, but opposite signs, of the OAM are split in frequency. In the conical geometry, this is formalized by the observation that the refractive indices $n_m(\omega,z)$ and $n_{-m}(\omega,z)$, solutions of Eq. \eqref{eq:SecularEquation}, are different.

At long wavelength, the frequencies of all modes with negative $m$ are close to the frequency $\omega_-$, see Eq. \eqref{eq:omega-}. 
Importantly, this holds true for any value of the waveguide radius, and in particular also for $R(z)\to0$. As seen in Sec. \ref{sec:cylinder}, for shrinking radius, the modes with positive $m$ tend to acquire a larger frequency. This suggests a strongly nonreciprocal behavior in the OAM within the range of frequencies  \eqref{eq:d}.

Figure \ref{fig:velocities} exemplifies the different behavior of two modes with the same absolute value of the OAM, but different sign. As discussed in Sec. \ref{sec:cone}, the $m=+2$ mode disappears at a given position along the cone, where its phase and group velocities become equal to the light velocity in the dielectric. This is, in fact, a general behaviour, as SPPs with positive $m$ are always radiated in the dielectric before reaching the tip. The radius corresponding to this condition defines a cutoff radius $R_m$. For modes with $|m|>1$, $R_m$ can be computed with a combination of analytical and numerical manipulations, by expanding  around the cutoff value $n\simeq\sqrt{\epsilon_d}+\Delta n $, $\Delta n\ll1$ \cite{Chang2007}. 
The resulting equation is provided in App. \ref{app:cutoff_radius}
and only has solutions for $m>1$.
For $m=1$ there is an effective cutoff in the sense that $n$ reaches $\sqrt{\epsilon_d}$ exponentially as $R\to0$, as in conventional metals \cite{Chang2006}.
\begin{figure}[h]
    \centering
    \includegraphics[width=1\columnwidth]{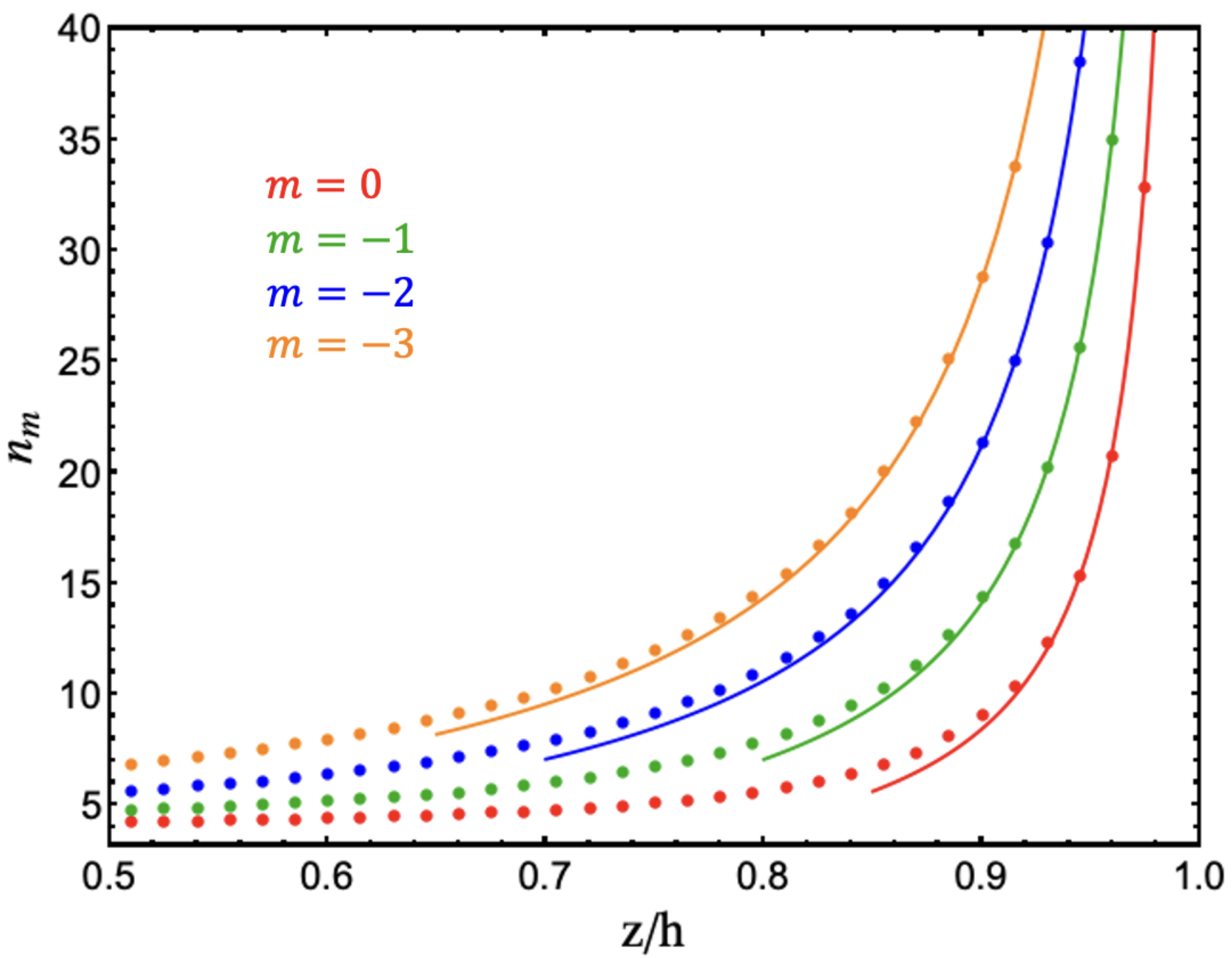}
    \caption{Refractive index of the $m=0,-1,-2,-3$ modes as a function 
    of the position along the axis of the cone. 
    The dots represent the numerical solutions of \eqref{eq:SecularEquation}, 
    while the solid lines are the best fits to  the expression \eqref{eq:nsmallR} with the single parameter $g_m$. 
    In this plot, $\omega/\omega_p=0.4$, $\epsilon_d=10$, $\epsilon_W=10$ and $\beta=10$.
    }
    \label{fig:enneminus3theo_w04}
\end{figure}

Conversely, modes with $m\le0$ always reach the end of the cone.
This statement comes from the analysis of the dispersion relations, as presented in Sec. \ref{sec:cylinder}. 
We can also support it using the considerations of Sec. \ref{sec:cone}, and in particular showing 
that $n$ never reaches the value $\sqrt{\epsilon_d}$. In fact, the numerical interpolation of 
the solutions of \eqref{eq:SecularEquation} shows that the refractive index diverges for $R\to0$ with the law
\begin{equation}\label{eq:nsmallR}
    n_m\simeq\frac{g_m}{k_0 R}\;.
\end{equation}
The coefficient $g_m$ can be computed in closed form for $m=0$ 
\begin{equation}\label{eq:g0}
    g_0=
    \sqrt{\frac{4\epsilon_d/\epsilon_W}{W(\exp\left\{2\gamma\right\}\epsilon_d/\epsilon_W)}}\,,
\end{equation}
where $W$ is the product logarithm function and $\gamma$ the Euler constant. Equation  \eqref{eq:g0} also follows in the same form for a conventional metal \cite{Stockman04}.
For $m\ne0$, we could not obtain analytical results for $g_m$, but the scaling law \eqref{eq:nsmallR} is confirmed by numerics to high accuracy, see Fig. \ref{fig:enneminus3theo_w04}. 
We conclude that a conical tip can support not only the $m=0$ mode until its end, but also all modes with $m<0$, an effect not achievable in a conventional metal. Inverting the direction of the vector $\mathbf{b}$, see Fig. \ref{fig:cone}, the set of SPP modes with opposite sign of the OAM is focused at the tip.

\subsection{Multiple Weyl nodes} \label{sec:multiple}

Finally, we present a more general implementation of the model, taking a well-studied material as an example. 
The theory presented in Sec. \ref{sec:Model} formally includes one pair of Weyl points and directly applies 
to materials like $\left(\mbox{Cr},\mbox{Bi}\right)_2\mbox{Te}_3$ \cite{Belopolski2025}. However, 
the experimentally more common situation is that multiple pairs of Weyl nodes are present. 
This can be modeled  by introducing a set of vectors $\mathbf{b}_j$, 
each connecting a negative- to a positive-chirality node. As the vectors generically point in different directions, 
the resulting permittivity acquires off-diagonal components from all of them. 
Nevertheless, the macroscopic observables from such components, eg., Faraday and Kerr rotations and the AHE, will be determined 
by the combined effect of all the Weyl nodes, i.e., by the sum $\sum_j\mathbf{b}_j$, 
which is independent of how the nodes are pairwise connected.\\
As shown in \cite{Weylguide2024} and Sec. \ref{sec:Model}, the SPP dispersion asymmetry and the ensuing 
OAM nanofocusing originate only from the antisymmetric part of the permittivity tensor, 
irrespective of its microscopic origin. This can be operatively determined by the measured (or otherwise computed) anomalous Hall conductivity of a WS \cite{Yang2011}. In our simplified model, only the universal DC limit is relevant, which can be interpreted as the anomalous Hall conductivity of an effective two-node model,  with the $\mathbf{b}$ vector given by the sum of all the pairwise node separations, see Eq. \eqref{eq:sigmaH}. 

As a case study, we consider the layered Kagome material $\mbox{Co}_3\mbox{Sn}_2\mbox{S}_2$ in its ferromagnetic phase, with a Curie temperature $T_C\sim175\,\mbox{K}$ \cite{Liu2018,Xu2018,Wang2018,Kanagaraj2022}. It hosts a topological phase with multiple pairs of Weyl nodes and associated Fermi arcs \cite{Liu2019,Morali2019,Kanagaraj2022}, which determines its characteristic magneto-optical responses \cite{Okamura2020}. Remarkably, samples of this material can be cleaved and reduced to nanoflakes \cite{Guguchia2020,Yang2020}.
We can therefore adapt the model of Sec. \ref{sec:nanofocusing} to a conical tip with $z$ axis perpendicular to the $\mbox{Co}$ planes and use the measured Hall conductivity 
$\sigma_{H}\approx1.1\times10^{3}/(\mbox{\ensuremath{\Omega}cm})$ \cite{Liu2018}, plasma frequency  
$\hbar\omega_{p}=258\,\mbox{meV}$ and dielectric constant $\epsilon_{W} \approx 25$ \cite{Xu2020}. 
From these data, we determine the parameters defined in Sec. \ref{sec:Model} 
as $\beta=16$ and $R_0=2c/\omega_p \approx 1530\,\mbox{nm}$. 

\begin{figure}
    \centering
    \includegraphics[width=0.47\linewidth]{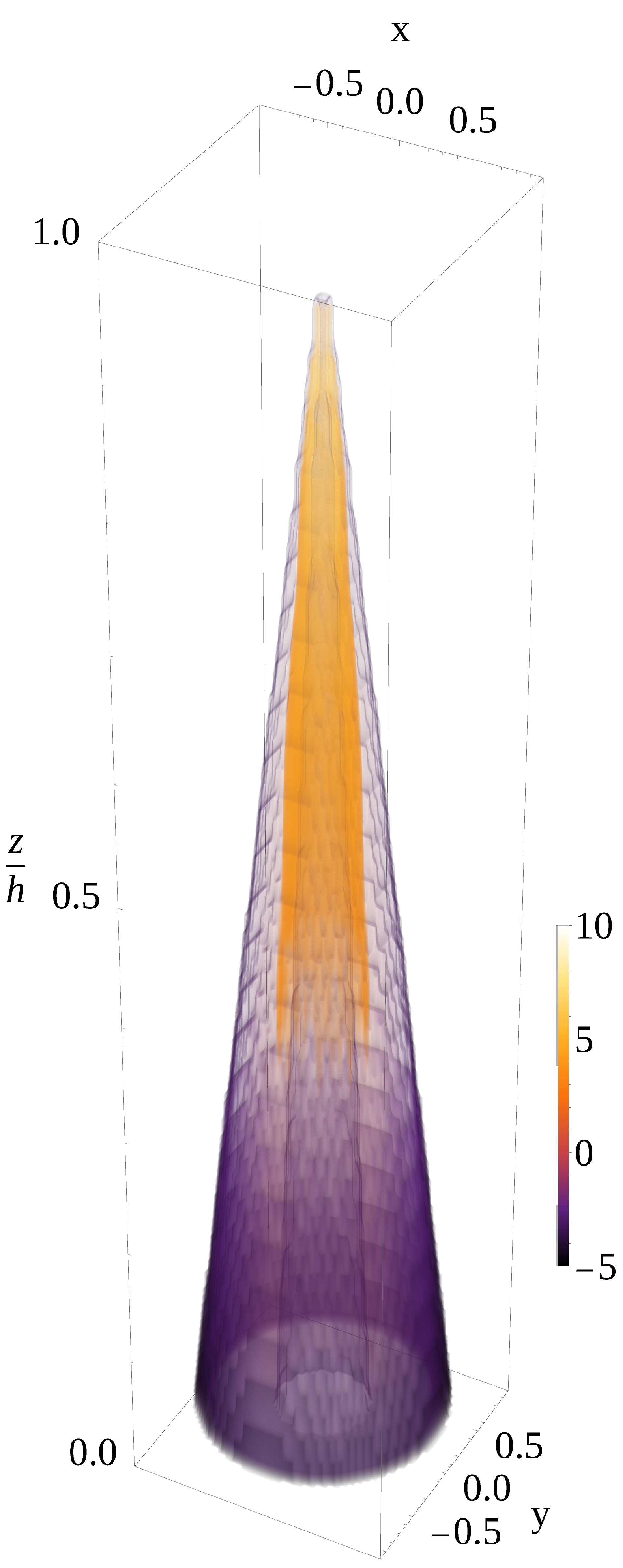}
    \includegraphics[width=0.47\linewidth]{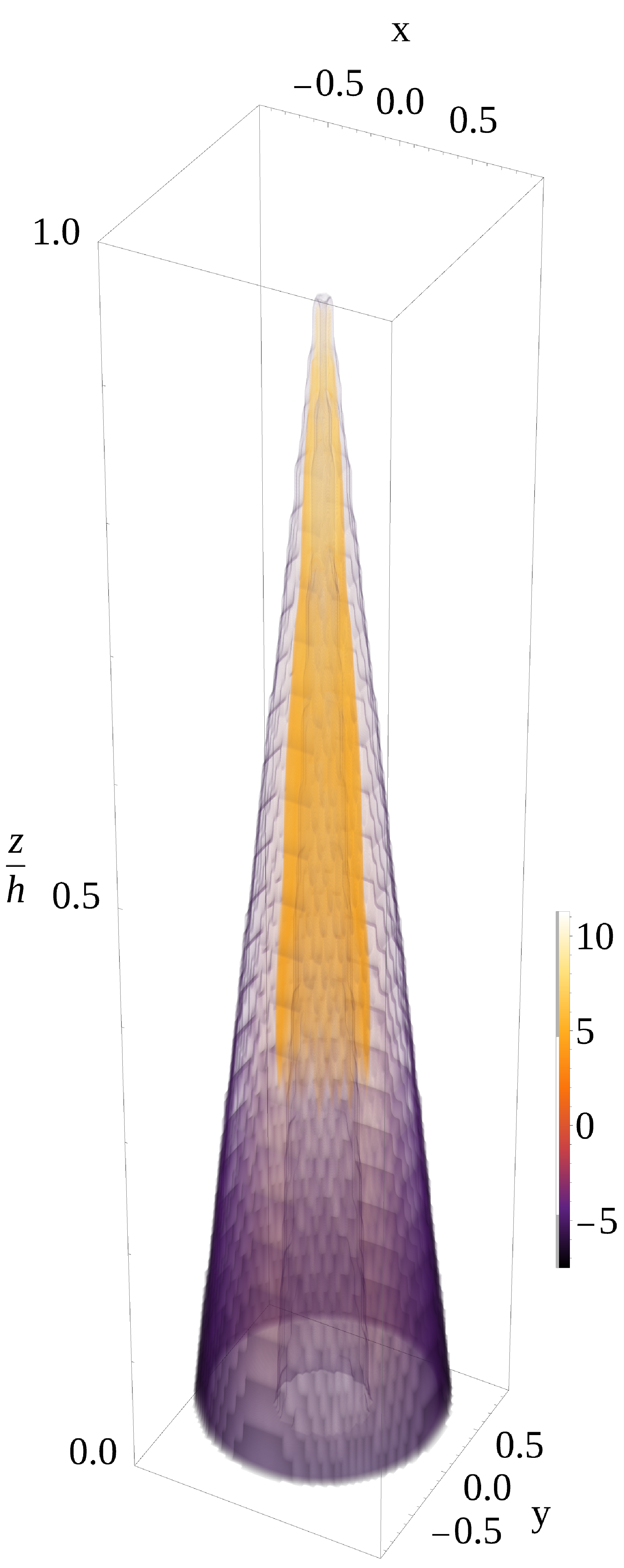}
    \caption{Logarithm of the LFIE Eq. \eqref{eq:LFIE} in the region $0.5R(z)<r<1.5R(z)$ around the surface of the cone, 
    for material parameters $\beta=16$, $\epsilon_W=25$ and frequency $\omega=0.75\omega_p$. 
    Lengths are in units of $c/\omega_p$. Left: $m=-1$, right $m=-2$.}
    \label{fig:intensity}
\end{figure}

We numerically verified the occurrence of the OAM nanofocusing for $m\le0$. In particular, in Fig. \ref{fig:intensity}, 
we show the LFIE \eqref{eq:LFIE} in logarithmic scale for the first two modes carrying nontrivial OAM, 
on a shell around the surface of the cone. The enhancement factor grows by several orders of magnitude at the end of the tip. 
We also provide numerical values for the cutoff radii for the lowest $m>0$ modes in Table \ref{tab:RcutoffCob}.

\begin{table}[h]
\centering
    \setlength{\tabcolsep}{10pt}
    \renewcommand{\arraystretch}{1.5}
    \begin{tabular}{|c|cccc|}
    \hline
    $m$ &  $+2$ &  $+3$ & $+4$ &  $+5$  \\ 
    \hline
    $R_{m}/R_0$ & $3.7$ & $7.4$ & $11.1$ & $14.6$ \\ 
    \hline
    \end{tabular} 
    \caption{Cutoff radius $R_m$ (in units of $R_0=2c/\omega_p$) of radiated modes at frequency $\omega=0.7\omega_p$, 
    for $\beta=16$, $\epsilon_W=25$, $\epsilon_d=1$. 
    The radii are obtained by finding the numerical roots of Eq.~\eqref{eq:RcutoffWeylCompact}.}
    \label{tab:RcutoffCob}
\end{table}

Finally, we note that our analysis accounts for the modes that have at least one component exponentially localized at the boundary \cite{Weylguide2024}. Nevertheless, also bulk modes can be present in the observed range of frequencies, which also can have a nonvanishing OAM number. Our results, e.g., the values in Table \ref{tab:RcutoffCob}, do not apply to these waveguide modes. These are linearly independent modes and should not affect the predicted behavior. In the presence of terms which couple SPP and waveguide modes, the different spatial localization is likely to make the hybridization matrix element small.

\section{Discussion}\label{sec:discussion}
Our study demonstrates that a conical tip made of a time-reversal broken WS supports multiple SPP modes, labeled by their OAM. In particular, within the frequency range specified by \eqref{eq:d}, all bands with a given sign of the OAM propagate to the cone apex, while all bands with the opposite sign are radiated before. We have shown this by analyzing the recently computed SPP bands of cylindrical waveguides, by numerically evaluating the field intensity enhancement, and by analytically estimating the cutoff radius of the radiated modes. Such OAM nanofocusing overcomes the limitations of conventional metals, in which only the $m=0$ mode can be focused at the tip.
The continuum description adopted in this work, underlying the Hamiltonian \eqref{eq:Hchi}, carries inherent limitations. First, it is a good approximation for the electron dynamics at scales much larger than the lattice constants. Moreover, it neglects the effect of finite size quantization and of the electron scattering at the boundaries \cite{Fuchs1938,Mayadas1970,Buccheri2022}, which may already play an important role in cylindrical nanowires with transverse size $\mathcal{O}(10\mbox{nm})$  \cite{DeMartino2021}.
Other limitations of our theory arise from neglecting dissipation effects and the wavevector dependence of the permittivity (local approximation). In principle, it is possible to retain the full functional form of the off-diagonal elements and modify the boundary conditions to take into account higher powers of the wavevector. While detailed theoretical and experimental estimates are available \cite{Zhou2018,Kanagaraj2022}, this comes at the cost of an increase in complication. We expect the predicted effect to be robust, as it relies on the splitting of SPP modes with different sign of the OAM, which was determined in the long wavelength limit \cite{Weylguide2024}. Even if the SPP dispersions are modified at larger wavevector, this should not affect the shift of the negative OAM modes toward the asymptote \eqref{eq:omega-}.
Inclusion of dissipation may affect the SPP dispersions. 
In related setups, it is known that nonreciprocity of SPP survives in some regimes \cite{Gangaraj2019}, but their determination requires the analysis of dissipation mechanisms, which is beyond the scope of this work.
All of the above are commonly implemented approximations \cite{Stockman04} and the prediction for metallic tips has been experimentally verified and applied with consistency \cite{Ropers2007,Shi2021}.

The field of topological materials has received much attention in recent years, one reason being that the topological properties associated with their nontrivial band structure are associated with specific magneto-optical responses \cite{Guo2023}. In this context, our work demonstrates that the chiral anomaly of topological WSs \cite{Jia2016}, which combine a nontrivial electrodynamics with a metallic permittivity, offers a route to manipulate the OAM degree of freedom down to the nanoscale.
The relevance of our study is both fundamental, as it unveils a new phenomenon characterizing a class of topological materials and allows for a deeper theoretical understanding of their phenomenology, and applied, in connection with other fields of physics. 
Indeed, the OAM nanofocusing should allow optical manipulation of particles with a much smaller size than presently achievable \cite{He1995,Gahagan1996,Friese1998,Galajda2002}. Despite being a single-photon property, the OAM needs to be transferred to the structure to be manipulated in order to set it in motion, which requires a sizeable matrix element, only achieved by a beam section comparable to the size of the target object.
 This makes it practically difficult to go below the $\mu$m scale, while the described effect would allow to set in motion much smaller structures, with the aid of a WS tip.
Many applications in the field of metrology are likely to benefit from the ability to narrowly focus twisted light. In addition to rotation spectroscopy \cite{Guo2020}, we point out near-field scanning optical microscopy (NSOM) \cite{Hecht2000} and tip-enhanced Raman spectroscopy (TERS) \cite{Deckert2017,Pienpinijtham2022}: versatile, non-destructive techniques which can achieve the high resolutions characteristic of tunneling microscopy and cantilever tips \cite{Garoli2014}. 
Adding OAM as a property of the incident radiation would potentially enhance these spectroscopic tools, e.g., selectively addressing different Raman transitions \cite{Brullot2016}.
Finally, OAM has emerged as a promising resource for quantum computation architectures, because of the large associated Hilbert space \cite{Zia2023}, robustness \cite{Meglinski2024} and versatility \cite{Lou2024}. The nontrivial optical properties of the topological WS phases would add to this field a way to control an important degree of freedom and to miniaturize architectures. 

\emph{Data availability:} All data underlying the figures in this paper are available at Zenodo, see \textcolor{blue}{https://doi.org/10.5281/zenodo.15772645}

\begin{acknowledgments}
We thank E. Di Fabrizio for inspiring and useful discussions. 
M.~P.~is founded through DM 118/2023 - Inv. 4.1, project "Light-matter interactions in topological semimetals", CUP E14D23001640006, PNRR. F.~B.~acknowledges financial support by NQSTI PE0000023 PNRR, through the project TOPMASQ, CUP E13C24001560001. 
R.~E.~acknowledges funding by the Deutsche Forschungsgemeinschaft (DFG, German Research Foundation) under Projektnummer 277101999 - TRR 183 (project A02) and under Germany's Excellence Strategy - Cluster of Excellence Matter and Light for Quantum Computing (ML4Q) EXC 2004/1 - 390534769.
\end{acknowledgments}

\appendix

\section{Field solutions}
\label{app:polarization_vectors}

Here we briefly review the solution of the electrodynamics equations
\begin{gather}
    \label{eq:GaussE}\boldsymbol{\nabla}\cdot\mathbf{D}=0, \\
    \label{eq:GaussB}\boldsymbol{\nabla}\cdot\mathbf{B}=0, \\
    \label{eq:Faraday}\boldsymbol{\nabla}\times\mathbf{E}-\mathrm{i}\omega\mathbf{B}=0, \\
    \label{eq:AmpMax}\boldsymbol{\nabla}\times\mathbf{H}+\mathrm{i}\omega\mathbf{D}=0,
\end{gather}
for a cylindrical waveguide \cite{Weylguide2024}.
The Weyl node separation enters in the constitutive equation $\mathbf{D}= \hat{\varepsilon}\mathbf{E}$  in the Weyl semimetal
via the permittivity tensor \cite{Zyuzin2012,Hofmann2016}, see Sec. \ref{sec:Model}. 
 After the factorization 

 \begin{equation}\label{eq:factorization}
     \mathbf{E}_m = \widetilde{\mathbf{E}}_m(r) e^{iq_zz} e^{im\varphi}\,, 
 \end{equation}
electric field in the WS takes the form 
 $\widetilde{\mathbf{E}}_m=\widetilde{\mathbf{E}}_m^++\widetilde{\mathbf{E}}_m^-$
 with
\begin{equation}
        \frac{\widetilde{\mathbf{E}}_m^\pm}{c}=\frac{\mathrm{i}a_\pm^{(m)}}{n}
	\left\{
        \begin{bmatrix}
            \frac{m}{\kappa_\pm r}I_m\left(\kappa_\pm r\right)  \\%
		  \mathrm{i}I_m'\left(\kappa_\pm r\right)\\
		  0
        \end{bmatrix}
		+\gamma_\pm
        \begin{bmatrix}
			I_m' \left(\kappa_\pm r\right) \\
			\frac{ \mathrm{i} m}{\kappa_\pm r} I_m\left(\kappa_\pm r\right)\\
			\frac{\mathrm{i}n k_0 \kappa_\pm}{\left(k_0^2\mathcal{E}+\kappa_\pm^2\right)}I_m\left(\kappa_\pm r\right)
		\end{bmatrix}
		\right\}
\end{equation}
while in the dielectric it reads
\begin{equation}
    \frac{\widetilde{\mathbf{E}}_m^d}{c}=
            \frac{\mathrm{i}d_1^{(m)}}{n}
        \begin{bmatrix}
            \frac{m}{\kappa_d r}K_{m}(\kappa_d r) \\
            \mathrm{i}K'_{m}(\kappa_d r) \\
            0
        \end{bmatrix}
        +
                \frac{\mathrm{i}d_2^{(m)}}{n}
        \begin{bmatrix}
            K'_{m}(\kappa_d r) \\
         \frac{\mathrm{i} m}{\kappa_d r}K_{m}(\kappa_d r) \\
            \frac{\mathrm{i}\kappa_d}{n k_0}K_{m}(\kappa_d r)
        \end{bmatrix} .
\end{equation}
Here, $\gamma_\pm$, $\kappa_\pm$, $\kappa_d$ are given in \eqref{eq:kappa_diel} 
and \eqref{eq:kappagamma} of the main text.
Similarly, the magnetic induction field can be presented as
\begin{equation}
    \widetilde{\mathbf{B}}_m=\left(\widetilde{\mathbf{B}}_m^++ \widetilde{\mathbf{B}}_m^-\right)
    \Theta\left(R-r\right)+\widetilde{\mathbf{B}}_m^d\Theta\left(r-R\right),
\end{equation}
with
\begin{equation}
        \widetilde{\mathbf{B}}_m^\pm=a_\pm^{(m)}
	\left\{
        \begin{bmatrix}
		I_m' \left(\kappa_\pm r\right) \\
		\frac{\mathrm{i} m}{\kappa_\pm r} I_m\left(\kappa_\pm r\right)\\
		\frac{\mathrm{i}\kappa_\pm}{n k_0}I_m\left(\kappa_\pm r\right)
		\end{bmatrix}
        +\frac{1}{\gamma_\pm}
        \begin{bmatrix}
            \frac{m}{\kappa_\pm r}I_m\left(\kappa_\pm r\right)  \\
		  \mathrm{i}I_m'\left(\kappa_\pm r\right)\\
		  0
        \end{bmatrix}
	\right\}
\end{equation}
in the Weyl semimetal and
\begin{equation}
    \widetilde{\mathbf{B}}_m^d=d_1^{(m)}
    \begin{bmatrix}
        K'_{m}(\kappa_d r) \\
    \frac{\mathrm{i} m}{\kappa_d r}K_{m}(\kappa_d r) \\
        \frac{\mathrm{i}\kappa_d}{n k_0}K_{m}(\kappa_d r)
    \end{bmatrix}
   + d_2^{(m)} \frac{\epsilon_d}{n^2}
    \begin{bmatrix}
        \frac{m}{\kappa_d r}K_{m}(\kappa_d r) \\
        \mathrm{i}K'_{m}(\kappa_d r) \\
        0
    \end{bmatrix}
\end{equation}
in the dielectric. The coefficients $a_\pm^{(m)}$, $d_{1,2}^{(m)}$ are determined by the boundary conditions, see sec. \ref{sec:cylinder}. 
\begin{figure}
    \centering
    \includegraphics[width=0.47\linewidth]{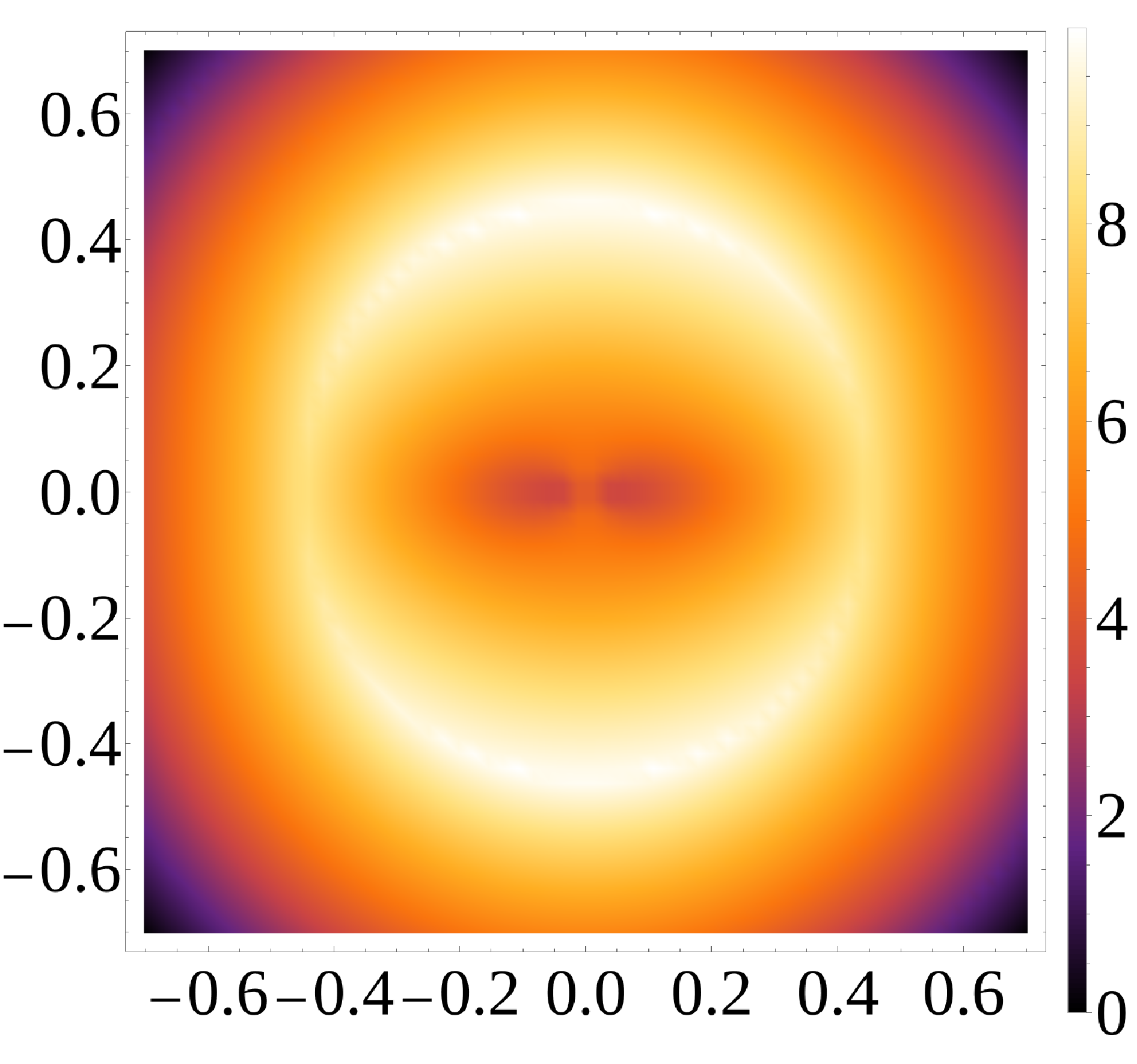}
    \includegraphics[width=0.51\linewidth]{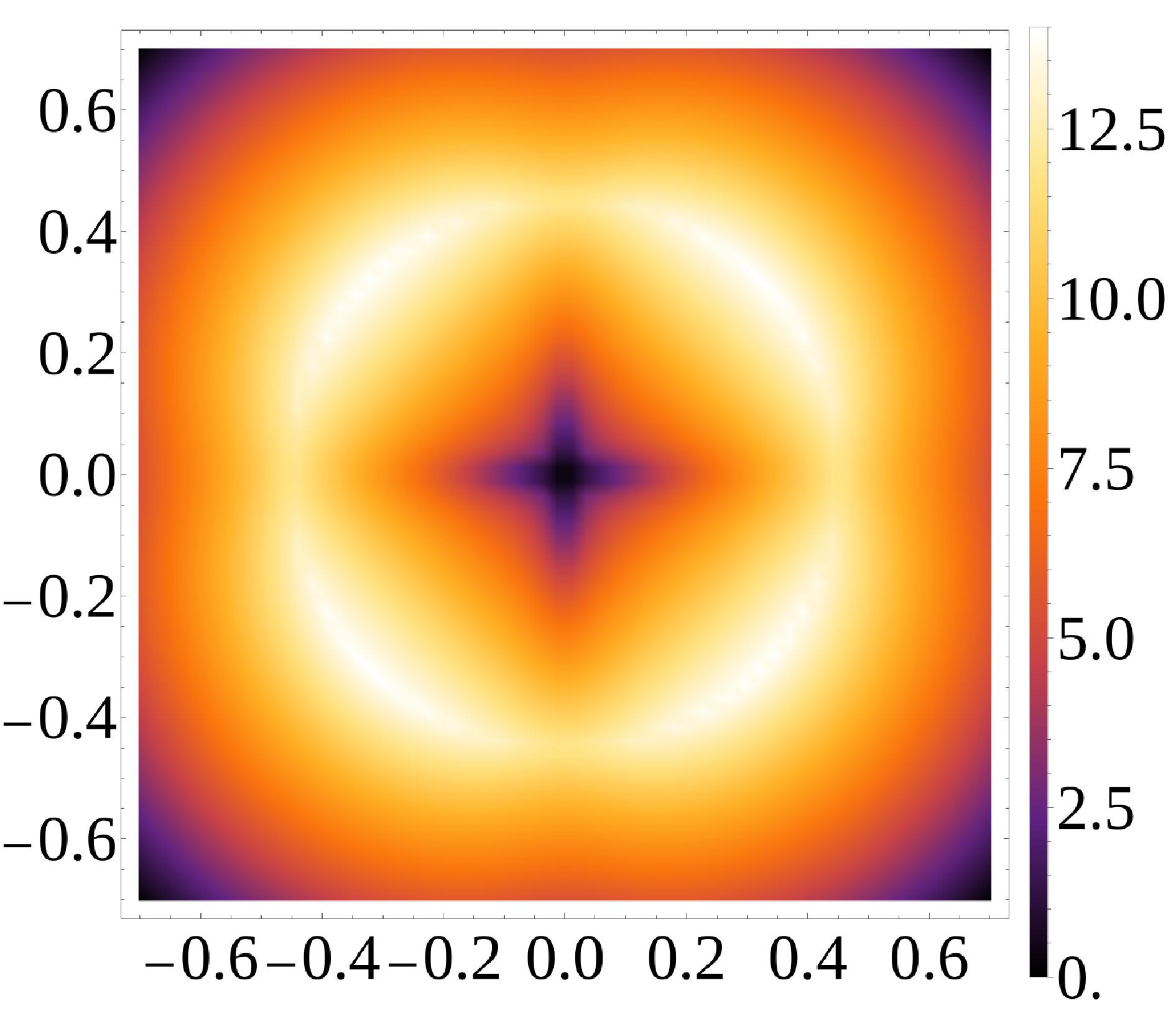}
    \caption{Illustration of the radial profile of the SPP modes. Here the logarithm of the LFIE Eq. \eqref{eq:LFIE} is shown, with parameters $\beta=16$, $\epsilon_W=25$, $\omega=0.75\omega_p$ and $m=-1$ (left), $m=-2$ (right). On the axes, the $x$ and $y$ coordinates are in units of $c/\omega_p$.}
    \label{fig:radial}
\end{figure}
As in the permittivity, we consistently retain the lowest order $\mathcal{O}\left(q^0\right)$ in the wavevector when doing so \cite{Langreth1989,Zhou2018}. This neglects the contribution of the surface charges and densities associated with the Fermi arcs, which appear at the following order. While this can affect the SPP dispersion, it should not instead alter the asymptote Eq. \eqref{eq:omega-}, which depends on the quasi-universal value of the AHE only. Examples of the radial profile of the SPP modes are shown in Fig. \ref{fig:radial} and exhibit clear surface localization.

In order to study a conical geometry, we follow \cite{Stockman04} and approximate the cone as a series of cylindrical waveguides with $z$-dependent radius $R(z)$, see Eq.~\eqref{zdepradius}. While translation invariance along $z$ is broken, under the hypothesis of a slowly varying radius, one can employ an Ansatz modeled on Eq. \eqref{eq:factorization}, namely, Eq. \eqref{eq:Epolvec} in the main text. The wavevector $q_z$ is replaced by $k_0 n$ using the relation \eqref{eq:vpvg}, with a position-dependent refractive index $n(\omega,z)$. The factorized plane wave $e^{iq_zz}$ is in turn replaced by a function in the form $e^{i\varphi(z)}$. Inserting this Ansatz into the wave equation, see App. A of \cite{Weylguide2024}, and assuming a slow variation of the amplitude $\mathcal{A}$ with respect to the phase, one obtains the consistency condition
\begin{equation}
    k_{0}^{2}n^2=-i\partial_{z}^{2}\varphi+\left(\partial_{z}\varphi\right)^{2}\;.
\end{equation}
Under the hypothesis
\begin{equation}\label{eq:dervarphi}
    \left|\partial_z^2\varphi\right|\ll \left(\partial_z\varphi\right)^2\;,
\end{equation}
it is possible to validate the Ansatz, with \mbox{$\varphi(z) = k_0\intop^zdz' n(z')$}. This is the 
form presented in Eq. \eqref{eq:Epolvec}. In terms of the adiabatic parameter \eqref{eq:delta}, 
the condition \eqref{eq:dervarphi} is formulated as $\delta_m\ll1$. 
Under the adiabatic approximation, the relations
\begin{equation}
    \partial_z\mathbf{E}_m\simeq \mathrm{i} \frac{n\omega}{c} \mathbf{E}_m\quad,\quad
    \partial_z^2\mathbf{E}_m\simeq- \frac{n^2\omega^2}{c^2}\mathbf{E}_m
\end{equation}
hold.

\section{Cutoff radius}\label{app:cutoff_radius}

The cutoff radius of the modes which do not reach the tip of the cone can be obtained from the secular equation of the matrix \eqref{eq:MatrixBoundCondMain} for $n\simeq\sqrt{\epsilon_d}+\Delta n$, with $\Delta n\ll1$ \cite{Chang2007}. This implies evaluating the determinant for small $v$, since
\begin{equation}
    v=R\kappa_d=\zeta \sqrt{\left(n+\sqrt{\epsilon_d}\right)\left(n-\sqrt{\epsilon_d}\right)}\to0,
\end{equation}
when $n\to\sqrt{\epsilon_d}$, where we use the notation $\zeta=Rk_0$. 
In doing this one has to use the fact that
    \begin{align}
    \frac{1}{n^2}&\simeq\frac{1}{\epsilon_d}-\frac{v^2}{\epsilon_d^2 {\zeta}^2}+\mathcal{O}(v^4)
\end{align}
and the behaviour of the Bessel functions $K_m$ for $v\to0$, for the cases $|m|=1$ and $|m|>1$.
\subsection{Case $|m|>1$}
When $|m|>1$ and $v\to0$, the expansion of the Bessel functions $K_m$ results in
\begin{equation}
    \frac{K'_m(v)}{K_m(v)}\simeq-\frac{|m|}{v}-\frac{v}{2\left(|m|-1\right)},
\end{equation}
and one can verify that, in this limit, the determinant of \eqref{eq:MatrixBoundCondMain} is not singular. 
Therefore, when $n=\sqrt{\epsilon_d}$, the cutoff radius $R_m=\zeta_m /k_0$
solves the equation
\begin{equation}\label{eq:RcutoffWeylCompact}
    \mathcal{F}^+_b\left(\zeta_m\right)-\mathcal{F}^-_b\left(\zeta_m\right)=0\;.
\end{equation}
The functions above are defined as
\begin{eqnarray}
    \mathcal{F}^\pm_b &=& u_\pm^d\frac{u_\mp^d}{\gamma_\mp^d\mathcal{E}}\left[\frac{\epsilon_d}{\left(|m|-1\right)}-\frac{|m|}{\zeta_m^2}\right] \nonumber \\
    &+&
    u_\pm^d\left(\operatorname{sign}{m}-\frac{1}{\gamma_\pm^d}\frac{\epsilon_d}{\mathcal{E}}\right)\bigg(\frac{\gamma_\mp^d}{u_\mp^d}\operatorname{sign}{m}+\mathcal{I}_m^\mp\bigg) \nonumber \\
    &+&
    u_\pm^d\left(1-\frac{\operatorname{sign}{m}}{\gamma_\pm^d}\frac{\epsilon_d}{\mathcal{E}}\right)\bigg(\frac{\mathcal{I}_m^\mp}{\gamma_\mp^d}+\frac{\operatorname{sign}{m}}{u_\mp^d}\bigg),
\end{eqnarray}
where we use the notation
\begin{align}
    & u_\pm^d=u_\pm\left(n=\sqrt{\epsilon_d}\right), &
    \gamma_\pm^d&=\gamma_\pm\left(n=\sqrt{\epsilon_d}\right), \\
    &\mathcal{I}_m^\pm=\frac{I'_m\left(u_\pm^d\right)}{I_m\left(u_\pm^d\right)}.
\end{align}
In $\mathcal{F}^\pm_b$, the combination $\gamma_\pm\operatorname{sign}{m}$, and therefore terms proportional to $b\operatorname{sign}{m}$, which are responsible for the asymmetry $\zeta_{-m}\ne \zeta_{m}$, are explicit. This implies that there are no solutions if $m<0$. 
When $b=0$, Eq. \eqref{eq:RcutoffWeylCompact} correctly reproduces the metallic limit \cite{Chang2007}, where
\begin{equation}
    \frac{\epsilon_d}{\left(|m|-1\right)}+\frac{\left(\epsilon_d+\mathcal{E}\right)}{\zeta_m^2\left(\epsilon_d-\mathcal{E}\right)}|m|+\frac{\epsilon_d+\mathcal{E}}{\zeta_m\sqrt{\epsilon_d-\mathcal{E}}}\mathcal{I}_m=0,
\end{equation}
with $\mathcal{I}_m=I'_m\left(\zeta_m\sqrt{\epsilon_d-\mathcal{E}}\right)/I_m\left(\zeta_m\sqrt{\epsilon_d-\mathcal{E}}\right)$, and any $|m|>1$ has a finite cutoff radius.
\subsection{Case $|m|=1$}
In the case $|m|=1$ one has to consider the logarithmic divergence of the function $K_m(v)$ when $v\to0$, namely
\begin{equation}
    \frac{K'_1(v)}{K_1(v)}\simeq-\frac{1}{v}+\left(\log{\frac{v}{2}}+\gamma\right)v.
\end{equation}
Considering $\displaystyle\left|\log{\frac{v}{2}}\right|\gg\gamma$, Eq. \eqref{eq:BoundCond} in this approximation, leads us to the condition
\begin{equation}\label{eq:RcutoffWeylCompact1}
    \mathcal{G}^+_b\left(\zeta_m\right)-\mathcal{G}^-_b\left(\zeta_m\right)=0,
\end{equation}
with
\begin{eqnarray}
    \mathcal{G}^\pm_b &=& -u_\pm^d\frac{u_\mp^d}{\gamma_\mp^d}\frac{\epsilon_d}{\mathcal{E}}\left(\frac{1}{\epsilon_d\zeta_m^2}+2\log{\frac{v}{2}}\right) \nonumber \\
    &+&
    u_\pm^d\left(\operatorname{sign}{m}-\frac{1}{\gamma_\pm^d}\frac{\epsilon_d}{\mathcal{E}}\right)\bigg(\frac{\gamma_\mp^d}{u_\mp^d}\operatorname{sign}{m}+\mathcal{I}_m^\mp\bigg) \nonumber \\
    &+&
    u_\pm^d\left(1-\frac{\operatorname{sign}{m}}{\gamma_\pm^d}\frac{\epsilon_d}{\mathcal{E}}\right)\bigg(\frac{\mathcal{I}_m^\mp}{\gamma_\mp^d}+\frac{\operatorname{sign}{m}}{u_\mp^d}\bigg),
\end{eqnarray}
The mode $m=1$ has an effective cutoff radius in the sense that $n$ reaches $\sqrt{\epsilon_d}$ exponentially when $R\to0$. This is the same as for a conventional metal \cite{Chang2007}. In fact, when $b=0$ and $R\to0$ one recovers the result
\begin{equation}
    \Delta n(\zeta_m)\simeq\frac{2}{\sqrt{\epsilon_d}\zeta_m^2}\exp{\left[\frac{2\left(\epsilon_d+\mathcal{E}\right)}{\zeta_m^2\epsilon_d\left(\epsilon_d-\mathcal{E}\right)}\right]},
\end{equation}
due to the logarithmic term in the equation \eqref{eq:RcutoffWeylCompact1}.

\bibliography{references}

\end{document}